\documentclass{mem}
\usepackage{graphicx}
\usepackage{txfonts}
\usepackage{natbib}
\usepackage[figuresright]{rotating}
\usepackage[a4paper,breaklinks,dvipdfm]{hyperref}
\usepackage{balance}
\idline{0}{1}
\begin{document}
\def\teff{$T\rm_{eff}$ }
\def\kms {\,$\mathrm{km\, s^{-1}}$ }
\def\kmss {\,$\mathrm{km\, s^{-1}}$}
\def\ms {$\mathrm{m\, s^{-1}}$ }

\newcommand{\Teff}{\ensuremath{T_\mathrm{eff}}}
\newcommand{\g}{\ensuremath{g}}
\newcommand{\gf}{\ensuremath{gf}}
\newcommand{\loggf}{\ensuremath{\log\gf}}
\newcommand{\glog}{\ensuremath{\log\g}}
\newcommand{\pun}[1]{\,#1}
\newcommand{\cobold}{\ensuremath{\mathrm{CO}^5\mathrm{BOLD}}}
\newcommand{\linfor}{Linfor3D}
\newcommand{\xx}{\ensuremath{\mathrm{1D}_{\mathrm{LHD}}}}
\newcommand{\punms}{\mbox{\rm\,m\,s$^{-1}$}}
\newcommand{\punkms}{\mbox{\rm\,km\,s$^{-1}$}}
\newcommand{\abuhe}{\mbox{Y}}
\newcommand{\grav}{\ensuremath{g}}
\newcommand{\mlp}{\ensuremath{\alpha_{\mathrm{MLT}}}}
\newcommand{\mlpcm}{\ensuremath{\alpha_{\mathrm{CMT}}}}
\newcommand{\moh}{\ensuremath{[\mathrm{M/H}]}}
\newcommand{\senv}{\ensuremath{\mathrm{s}_{\mathrm{env}}}}
\newcommand{\shelio}{\ensuremath{\mathrm{s}_{\mathrm{helio}}}}
\newcommand{\smin}{\ensuremath{\mathrm{s}_{\mathrm{min}}}}
\newcommand{\spun}{\ensuremath{\mathrm{s}_0}}
\newcommand{\sstar}{\ensuremath{\mathrm{s}^\ast}}
\newcommand{\tauross}{\ensuremath{\tau_{\mathrm{ross}}}}
\newcommand{\ttaurelation}{\mbox{T$(\tau$)-relation}}
\newcommand{\Ysurf}{\ensuremath{\mathrm{Y}_{\mathrm{surf}}}}
\newcommand{\mD}{\ensuremath{\left\langle\mathrm{3D}\right\rangle}}

\newcommand{\draftflag}{false}

\newcommand{\beq}{\begin{equation}}
\newcommand{\eeq}{\end{equation}}
\newcommand{\pdx}[2]{\frac{\partial #1}{\partial #2}}
\newcommand{\pdf}[2]{\frac{\partial}{\partial #2}\left( #1 \right)}

\newcommand{\var}[1]{{\ensuremath{\sigma^2_{#1}}}}
\newcommand{\sig}[1]{{\ensuremath{\sigma_{#1}}}}
\newcommand{\cov}[2]{{\ensuremath{\mathrm{C}\left[#1,#2\right]}}}
\newcommand{\xtmean}[1]{\ensuremath{\left\langle #1\right\rangle}}

\newcommand{\eref}[1]{\mbox{(\ref{#1})}}

\newcommand{\Vact}{\ensuremath{\nabla}}
\newcommand{\Vad}{\ensuremath{\nabla_{\mathrm{ad}}}}
\newcommand{\Veddy}{\ensuremath{\nabla_{\mathrm{e}}}}
\newcommand{\Vrad}{\ensuremath{\nabla_{\mathrm{rad}}}}
\newcommand{\Vraddiff}{\ensuremath{\nabla_{\mathrm{rad,diff}}}}
\newcommand{\cp}{\ensuremath{c_{\mathrm{p}}}}
\newcommand{\taueddy}{\ensuremath{\tau_{\mathrm{e}}}}
\newcommand{\vconv}{\ensuremath{v_{\mathrm{c}}}}
\newcommand{\Fconv}{\ensuremath{F_{\mathrm{c}}}}
\newcommand{\lmix}{\ensuremath{\Lambda}}
\newcommand{\Hp}{\ensuremath{H_{\mathrm{P}}}}
\newcommand{\Hptop}{\ensuremath{H_{\mathrm{P,top}}}}
\newcommand{\COBOLD}{{\sc CO$^5$BOLD}}

\newcommand{\changed}{}

\newcommand{\I}{\ensuremath{I}}
\newcommand{\Irot}{\ensuremath{\tilde{I}}}
\newcommand{\F}{\ensuremath{F}}
\newcommand{\Frot}{\ensuremath{\tilde{F}}}
\newcommand{\vsini}{\ensuremath{V\sin(i)}}
\newcommand{\vvsini}{\ensuremath{V^2\sin^2(i)}}
\newcommand{\vsinimu}{\ensuremath{\tilde{v}}}
\newcommand{\rotint}{\ensuremath{\int^{+\vsinimu}_{-\vsinimu}\!\!d\xi\,}}
\newcommand{\imu}{\ensuremath{m}}
\newcommand{\imupone}{\ensuremath{{m+1}}}
\newcommand{\nmu}{\ensuremath{N_\mu}}
\newcommand{\msum}[1]{\ensuremath{\sum_{#1=1}^{\nmu}}}
\newcommand{\wmu}{\ensuremath{w_\imu}}

\newcommand{\tchar}{\ensuremath{t_\mathrm{c}}}
\newcommand{\Nt}{\ensuremath{N_\mathrm{t}}}

\title{Solar abundances and granulation effects}

\author{
E. Caffau     \inst{1}\and
H.-G. Ludwig  \inst{2,1}\and
M. Steffen   \inst{3}
}

\institute{
GEPI, Observatoire de Paris, CNRS, Universit\'e Paris Diderot, Place
Jules Janssen, 92190
Meudon, France
\and
CIFIST Marie Curie Excellence Team
\and
Astrophysikalisches Institut Potsdam, An der Sternwarte 16, D-14482 Potsdam, Germany
}
\authorrunning{Caffau et al.}
\titlerunning{Solar abundances and granulation effects}
\offprints{E. Caffau}

\abstract
{
The solar abundances have undergone a major downward revision
in the last decade, reputedly as a result of employing
3D hydrodynamical simulations to model the inhomogeneous structure
of the solar photosphere.
The very low oxygen abundance advocated by \citet{asplund04}, A(O)=8.66,
together with the downward revision of the carbon and nitrogen abundances,
has created serious problems for solar models to explain the helioseismic
measurements.

In an effort to contribute to the dispute we have re-derived
photospheric abundances of several elements
independently of previous analysis.
We applied a state-of-the art 3D (CO5BOLD) hydrodynamical simulation
of the solar granulation as well as different 1D model atmospheres
for the line by line spectroscopic abundance determinations. The
analysis is based on both standard disc-centre and disc-integrated spectral
atlases; for oxygen we acquired in addition spectra at different
heliocentric angles. The derived abundances are the result of equivalent
width and/or line profile fitting of the available atomic lines.
We discuss the different granulation effects on solar abundances
and compare our results with previous investigations.
According to our investigations hydrodynamical models are important
in the solar abundance determination, but are not responsible for the
recent downward revision in the literature of the solar metallicity.
}
\maketitle


\begin{table*}
\caption{Averaged 3D corrections in the case of $\xi _{\rm micro} =1.0$\kms and \mlp $=1.0$.}
\label{3dcor}
\begin{center}
{\scriptsize
\begin{tabular}{lrlllllll}
\hline
\noalign{\smallskip}
EL & N  & ion & Spec & \cobold\ & 3D-\xx & 3D-\mD & role of 3D & Ref.\\
   &    &     &      &          & [dex]  & [dex]  &            &     \\
\hline
\noalign{\smallskip}
Li & 1  &  \ion{Li}{i} & I/F & $1.03\pm 0.03$ &        &        & 3D-NLTE         &  \\
C  & 43 &  \ion{C}{i}  & I/F & $8.50\pm 0.06$ &  +0.02 & --0.03 & $\xi_{\rm micro}$           &  \\
N  & 12 &  \ion{N}{i}  & I   & $7.86\pm 0.12$ & --0.05 & --0.01 &  \mlp\                      & \citet{azoto} \\
O  & 10 &  \ion{O}{i}  & I/F & $8.76\pm 0.07$ &  +0.05 &  +0.01 & $\xi_{\rm micro}$           & \citet{oxy} \\
P  & 5  &  \ion{P}{i}  & I/F & $5.46\pm 0.04$ &  +0.03 &  +0.01 &                 & \citet{phos} \\
S  & 6  & \ion{S}{i}   & F   & $7.16\pm 0.05$ &  +0.04 &  +0.01 & $\xi_{\rm micro}$ & \citet{smult3}\\
Fe & 38 &  \ion{Fe}{i} & I   & $7.45\pm 0.06$ &  +0.11 &  +0.03 & $\xi_{\rm micro}$           &  \\
Fe & 15 &  \ion{Fe}{ii}& I/F & $7.52\pm 0.06$ &  +0.08 &  +0.05 & $\xi_{\rm micro}$           &  \\
Eu & 5  &  \ion{Eu}{ii}& I/F & $0.52\pm 0.03$ &  +0.01 &  +0.02 &                 & \citet{mucciarelli08} \\
Hf & 4  &  \ion{Hf}{ii}& I/F & $0.87\pm 0.04$ &  +0.02 &  +0.01 &                 & \citet{thhf} \\
Th & 1  &  \ion{Th}{ii}& I/F & $0.08\pm 0.03$ & --0.10 &        & Line asymmetry  & \citet{thhf} \\
\noalign{\smallskip}
\hline
\end{tabular}
}
\end{center}
\end{table*}

\section{Introduction}

In this work we would like to face the most common questions we are
confronted with in the analysis of the photospheric solar abundances: 
\begin{itemize}
\item ``Are 3D models important in the abundances determination?''
\item ``Are 3D models responsible for the downward revision of the solar metallicity?''
\end{itemize}
Our answer to the first question is yes.
As we know from previous investigations \citep{zolfito},
3D solar metallicity models do not experience the over-cooling in the 
external layers, not detected in 1D models, that metal-poor 3D models show.
One could then expect that 3D models are not fundamental for the solar abundance
determinations. If for some elements (P, Eu, Hf) the granulation effects are in fact
negligible, this is not the case for others, such as Fe, Th, and  also oxygen.
On top of that one should not forget that 1D models require some input parameters
(mixing-length parameter, \mlp, and microturbulence, $\xi_{\rm micro}$) 
implicit in 3D models; so that in the case such parameters are fundamental 
(\mlp\ for C, O, and Fe, $\xi_{\rm micro}$ for N)
3D models should be highly preferred.

\begin{table}
\caption{Influence of the microturbulence on the 3D corrections.}
\label{microt}
{
\begin{tabular}{ccc}
\hline
\noalign{\smallskip}
$\xi _{\rm \xx}$ & \multicolumn{2}{c}{$\left({\rm A(Y)_{\rm 3D}}-{\rm A(Y)_{\rm \xx}}\right)$ [dex]} \\
\kms\ & Flux & Intensity \\
\noalign{\smallskip}
\hline
\noalign{\smallskip}
0.6 & +0.036 &  - -\\
0.9 &  - -   & +0.037\\
1.2 & +0.130 & - - \\
1.5 &  - -   & +0.139\\
\noalign{\smallskip}
\hline
\end{tabular}
}
\end{table}

\begin{table*}
\caption{Photospheric solar abundances.}
\label{sunabbo}
\begin{center}
{
\begin{tabular}{lrlllll}
\hline
\noalign{\smallskip}
EL & N  & \cobold\ & AG89 & GS98 & AGS05 & AGSS09\\
\noalign{\smallskip}
\hline
\noalign{\smallskip}
Li & 1  & $1.03\pm 0.03$ & $1.16\pm 0.10$ & $1.10\pm 0.10$        & $1.05\pm 0.10$        & $1.05\pm 0.10$ \\
C  & 43 & $8.50\pm 0.06$ & $8.56\pm 0.04$ & $8.52\pm 0.06$        & $8.39\pm 0.05$        & $8.43\pm 0.05$ \\
N  & 12 & $7.86\pm 0.12$ & $8.05\pm 0.04$ & $7.92\pm 0.06$        & $7.78\pm 0.06$        & $7.83\pm 0.05$ \\
O  & 10 & $8.76\pm 0.07$ & $8.93\pm 0.035$& $8.83\pm 0.06$        & $8.66\pm 0.05$        & $8.69\pm 0.05$ \\
P  & 5  & $5.46\pm 0.04$ & $5.45\pm 0.04$ & $5.45\pm 0.04$        & $5.36\pm 0.04$        & $5.41\pm 0.03$ \\
S  & 9  & $7.16\pm 0.05$ & $7.21\pm 0.06$ & $7.33\pm 0.11$        & $7.14\pm 0.05$        & $7.12\pm 0.03$ \\
K  & 6  & $5.11\pm 0.09$ & $5.12\pm 0.13$ & $5.12\pm 0.13$        & $5.08\pm 0.07$        & $5.03\pm 0.09$ \\
Fe & 15 & $7.52\pm 0.06$ & $7.67\pm 0.03$ & $7.50\pm 0.05$        & $7.45\pm 0.05$        & $7.50\pm 0.04$ \\
Eu & 5  & $0.52\pm 0.03$ & $0.51\pm 0.08$ & $0.51\pm 0.08$        & $0.52\pm 0.06$        & $0.52\pm 0.04$ \\
Hf & 4  & $0.87\pm 0.04$ & $0.88\pm 0.08$ & $0.88\pm 0.08$        & $0.88\pm 0.08$        & $0.85\pm 0.04$ \\
Os & 3  & $1.36\pm 0.19$ & $1.45\pm 0.10$ & $1.45\pm 0.10$        & $1.45\pm 0.10$        & $1.40\pm 0.08$ \\
Th & 1  & $0.08\pm 0.03$ & $0.12\pm 0.06$ & ${\it 0.09\pm 0.02}$  & ${\it 0.06\pm 0.05}$  & $0.02\pm 0.10$ \\
   &    &                &                &                       &                       & \\
Z  &    & 0.0153         & 0.0189         & 0.0171                & 0.0122                & 0.0134 \\
Z/X&    & 0.0209         & 0.0267         & 0.0234                & 0.0165                & 0.0183 \\

\noalign{\smallskip}
\hline
\end{tabular}
}
\end{center}
Note: AG89=\citet{anders89}, GS98=\citet{grevesse98}, AGS05=\citet{sunabboasp}, and AGSS09=\citet{asplund09}.
\end{table*}


Our answer to the second question is no. We would like to remark
that this answer depends on the 1D reference model one is considering.
For all the elements, except N and Th, our 3D model gives an abundance larger than
the 1D model. Th is an exception because its lower 3D abundance is related
to the fact that the line is on the red wing of a stronger Fe-Ni blend, and
the asymmetry of this blend, correctly taken into account in 3D analysis,
lowers the Th contribution \citep{thhf}.
For N all the lines considered are of high excitation energy, so their
1D abundance is very sensitive on the choice of \mlp.
To try to catch the effects of granulation on the abundances determination,
we selected 1D models sharing the micro-physics with our 3D solar model
atmosphere.
But the comparison of abundances derived from 1D and 3D models is not unambiguously defined,
relying on the choice of \mlp\ and  $\xi_{\rm micro}$.


\section{Model atmospheres and observed data}

Our photospheric solar abundance analysis was performed by using
3D-\cobold\ solar models \citep{freytag02,freytag03,wedemeyer04}. 
The results here reported mostly rely on a solar
model covering 1.2\,h of solar time, 
with a box size of $5.6\times 5.6\times 2.3~{\rm Mm}^3$, 
a resolution of $140\times 140\times 150$,
12 opacity bins based on opacities stemming 
from the MARCS stellar atmosphere package \citep{marcsa,marcsb}. 
The model spans a range in optical depth of about $-6.7<\log\tau_\mathrm{Ross}<5.5$.
For details see \citet{oxy}.

We are here interested in the granulation effect on abundances,
and for this reason we selected two 1D models that share the micro-physics with
\cobold:
\begin{itemize}
\item \mD, a 1D model obtained by temporal and horizontal average 
over surfaces of equal (Rosseland) optical depth of the 3D model;
\item \xx, a 1D, plane parallel model, which employs the same micro-physics as \cobold.
\end{itemize}

For the photospheric solar abundance determinations,
as observed spectra, we considered the data
described in \citet{oxy} and spectra of solar intensity spectra for nine 
heliocentric angles observed at Kitt Peak with the McMath-Pierce Solar Telescope by W.~Livingston.


\section{3D corrections and solar abundances}

As we did in our previous work \citep{zolfito}, we define two 3D corrections as:
${\rm A(Y)_{\rm 3D}}-{\rm A(Y)_{\rm \xx}}$
${\rm A(Y)_{\rm 3D}}-{\rm A(Y)_{\rm \mD}}$
with Y the generic element and ${\rm A(Y)}=\log{n_{\rm Y}\over n_{\rm H}}+12$.
The first 3D correction takes into consideration 
the effects of convection on the 3D temperature structure,
the latter one the effects of fluctuations around the mean stratification.

For the elements so far analysed we can say that the effects of granulation on the
abundance analysis are not very large, but for the Sun, due to the high quality
spectra and to the high precision request in the abundance determination, 
they are not negligible. In Table\,\ref{3dcor} the average, over all lines as well as
disc-centre and disc-integrated when available, values are given.
For all the elements considered, the granulation effects due to
fluctuations around the mean stratification are small, being the highest of +0.05
for \ion{Fe}{ii}.
${\rm A(Y)_{\rm 3D}}-{\rm A(Y)_{\rm \xx}}$ on average is larger in absolute value,
and can be as large as 0.1\,dex (see iron).

Both 3D corrections are function of the microturbulence of the 1D models, and the first 
one of the mixing-length parameter as well.

\begin{figure}
\resizebox{\hsize}{!}{\includegraphics[clip=true,angle=0]{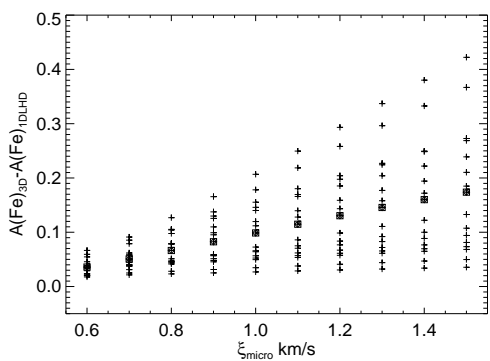}}
\caption{The 3D correction as a function of micro-turbulence, for 15 \ion{Fe}{ii} lines (crosses).
Squares depict the average over the 15 lines.}
\label{microcor}
\end{figure}

We changed the microturbulence in a reasonable range for disc-centre and
disc-integrated for 15 \ion{Fe}{ii} lines with 0.8\,pm$<$EW$<$8.8\,pm. 
The results are in Table\,\ref{microt}.
In Fig.\,\ref{microcor} this behaviour is shown, for each \ion{Fe}{ii} line, 
in the disc-integrated case.

The choice of the mixing-length parameter for the \xx\ model 
influences mainly the lines formed
deep in the photosphere, meaning atomic lines of high lower level energy.
We considered 12 \ion{N}{i} lines with $10.3~{\rm eV}<{\rm E_{\rm low}}<11.8~{\rm eV}$,
and changing \mlp\ of $^{+1.0}_{-0.5}$ 
with respect to the reference value of 1.0, we obtained
changes on ${\rm A(N)}_{\rm \xx}$ of $^{-0.09}_{+0.05}~{\rm dex}$,
which translate in analogous changes in the 3D correction.

The solar abundances we derived with the \cobold\ model are listed in 
Table\,\ref{sunabbo} and compared to other solar abundances compilations.
The majority of the disagreement one can see in the table are due to
improvements in the atomic data.
The solar metallicity has been computed using the  \cobold\ bases abundances,
when avaliable, and the solar abundances
from \citet{Lodders} for all other elements.


\section{Conclusions}

In the light of our work we think that the use of 3D models in the solar 
abundance determination is useful. This is for the following reasons:
\begin{itemize}
\item no constraint is necessary on \mlp\ and $\xi_{\rm micro}$;
\item a difference of few hundreds of dex in the abundance, negligible for
the majority of the stellar analysis, is important in the solar context. 
\end{itemize}


\balance 

\begin{acknowledgements}
  The authors E.C. and H.-G.L. acknowledge financial
  support from EU contract MEXT-CT-2004-014265 (CIFIST).
\end{acknowledgements}

\bibliographystyle{aa}

\end{document}